\documentclass[twocolumn,showpacs,preprintnumbers,amsmath,amssymb]{revtex4}

\usepackage[utf8]{inputenc}
\usepackage[dvips]{graphicx}
\usepackage{dcolumn}
\usepackage{bm}

\begin{document}

\title{Decay of $^{99m}$Tc as part of calibration procedure for $^{192}$Ir activity measurements using a well type ionization chamber}

\author{Marco A. Ridenti $^{1}$}

\author{Paulo R. Pascholati$^{1}$}

\author{Suzana Botelho$^{2,3}$}
\email{sbotelho@ipen.br}

\author{Josemary A. C. Gon\c{c}alves$^{1,2}$}

\author{Carmen C. Bueno$^{2,3}$}

\affiliation{$^{1}$Laborat\'orio do Acelerador Linear - Instituto de F\'isica - IFUSP,\\
CP 66318, 05315-970, S\~ao Paulo, Brazil.}

\affiliation{$^{2}$Instituto de Pesquisas Energ\'eticas e Nucleares - 
IPEN-CNEN/SP\\
 CP 11409, 05422-970, S\~ao Paulo, Brazil} 

\affiliation{$^{3}$Departamento de F\'isica - Pontif\'icia Universidade Cat\'olica de 
S\~ao Paulo - PUC/SP,\\ R. Marqu\^es de Paranagu\'a 111, 01303-050, Brazil}

\date{\today}

\begin{abstract}
In this work the linearity in the response of an ionization-chamber as a function of of $^{99m}$Tc activity was checked as part of a calibration procedure for $^{192}$Ir activity measurements. The non-linearity was found to be caused mainly by the electrometer range switching. A model based on the hypothesis that the non-linear effect did not influence the decay constant was tested. This method allowed to correct the non-linear effect and obtain the $^{99m}$Tc activity with good acuracy. 
\end{abstract}

\maketitle

\section{Introduction}

In this work the ionization-chamber current from a sample of a saline solution of $^{99m}$Tc was measured repeatedly, following its decay, in order to check the linearity of a well-type  ionization-chamber as part of a calibration procedure for $^{192}$Ir activity measurements. 

An ionization chamber used to perform activity measurements must have good linearity in its response as a function of the source activity within the required range of measured activity. There are several effects which could influence the linearity of an ionization-chamber, but the most important are usually saturation effects and the non-linearity of the current-measuring electronics. For instance, range switching is an important non-linearity effect of the current-measuring electronics \cite{Schrader}. In fact the results of measurements showed a non-linearity which is tipically associated with range switching, even though within the same range good linearity was found. We will also discuss a method to estimate the $^{99m}$Tc half-life using sets of data from two different ranges together in the same least squares fit of the logarithm of the measured current as a function of time. 

\section{Experimental Setup}

The experimental setup was described in details in previous work \cite{Goncalves}. The measurements were carried out using a well-type 4$\pi \, \gamma$ ionization chamber connected to a Keithley model 617 Programmable Electrometer, fully computer-controlled by means of specially designed software written in C, which allowed automatic current readings every 20 min, over a 3-min time bin. The input voltage was kept constant at 600 V in all measurements.

\section{Results}

\begin{figure}
\includegraphics[height=4.5cm, width = 8.7cm]{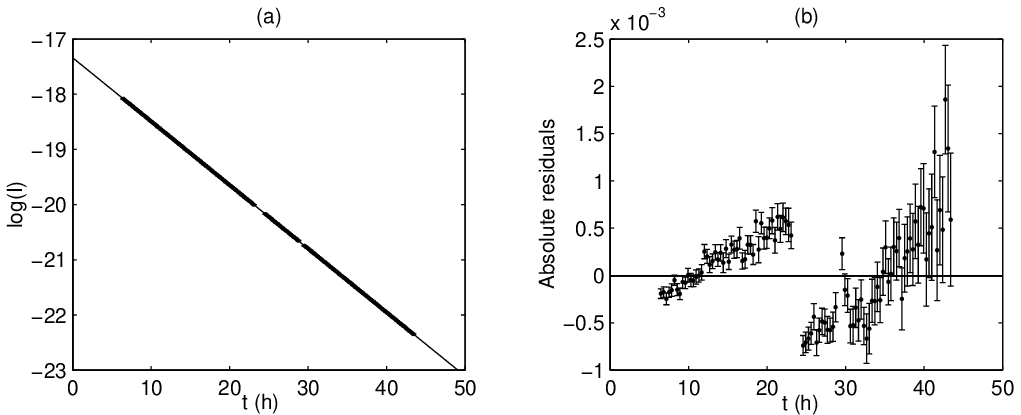}
\caption{\label{fig:graph1}  Simple linear least square fit of the logarithm of the ionization current as a function of time based on the fundamental law of exponential radioactive decay and residuals of the experimental points and the fitted values.}
\end{figure}

The data was fitted supposing that the fundamental law of exponential radioactive decay was valid, which was to expected if the ionization-chamber had good linearity.  Residuals analysis showed graphically the presence of a step discontinuity when the range switched from 20 nA to 2 nA (see figure \ref{fig:graph1}(b)). The residuals patern sugested that the set of data points changes systematically when the range switching occurs, in such a way that only the fitted linear coeficient would be different for the two different sets of data. Considering the data set of current measurements $I_{i}$, $i=1,2,...,N$, and the two subsets $y_{j}$, $i=1,2,...,M-1$ and $y_{k}$, $k=M,...,N$ ($M<N$) that correspond to the two different ranges, then this assumption can be expressed mathematically as follows

\begin{equation}
y_{i} = \log(I_{i}) = \left\{ \begin{array}{ll}
 -\lambda\,t_{i} + \alpha_{1} & \textrm{if $1 \leq i < M $}\\
 -\lambda\,t_{i} + \alpha_{2}  & \textrm{if $M \leq i < N $}\\
  \end{array} \right.
\end{equation}
where $log(I_{i})$ is the natural logarithm of the current $I_{i}$ and $\lambda$ is the \emph{desintegration constant}. In matrix formalism, this equation may be written as follows 

\begin{equation} 
\underbrace{\left( \begin{array}{c}
y_{1} \\
\vdots \\
y_{M-1} \\
y_{M} \\
\vdots \\
y_{N} 
\end{array}
\right)_{N\textrm{x}1}}_{\mathbf{Y}}
 =
\underbrace{\left( \begin{array}{ccc}
1 & 0 & t_{1} \\
\vdots & \vdots & \vdots \\
1 & 0 & t_{M-1} \\
0 & 1 & t_{M} \\
\vdots & \vdots & \vdots \\
0 & 1 & t_{N} 
\end{array} \right)_{N\textrm{x}3}}_{\mathbf{X}}
\hspace{-5mm}
\underbrace{\left( \begin{array}{c}
\alpha_{1} \\
\alpha_{2} \\
-\lambda 
\end{array}
\right)}_{\mathbf{A}}
\end{equation}
The fitted parameter matrix $\mathbf{\tilde{A}}$ is then given by the least square method in matrix formalism as 

\begin{equation}   
\mathbf{\tilde{A} = (X^{T} \, W^{-1} \, X)^{-1} \, X^{T} \, W^{-1} Y}
\end{equation}
where $\mathbf{W}$ is the covariance matrix of the data, which is given by
\begin{equation} 
\mathbf{W} 
 =
\left( \begin{array}{cccc}
\sigma_{y_{1}}^{2} & 0 & \vdots & 0 \\
0 & \sigma_{y_{2}}^{2} & \vdots & 0\\
\vdots & \vdots & \ddots & \ldots \\
0 & 0 & \ldots & \sigma_{y_{N}}^{2}
\end{array} \right) + 
\left( \begin{array}{ccc}
\scriptstyle \mathrm{cov}(y_{1}y_{2}) & \vdots & \scriptstyle \mathrm{cov}(y_{1}y_{N}) \\
\vdots & \ddots & \ldots  \\
 \scriptstyle \mathrm{cov}(y_{N}y_{1}) & \vdots & \scriptstyle \mathrm{cov}(y_{N}y_{N})  \\
\end{array} \right)
\end{equation}
where $\sigma_{y_{i}}$ is the standard deviation of $y_{i}$ and
\begin{equation}
\mathrm{cov}(y_{i}y_{j}) = \left\{ \begin{array}{ll}
 0 \, , & \textrm{if $i < M  \land j \geq M  \lor j < M  \land i \geq M $ } \\
 \delta_{i} \, \delta_{j} \, , & \textrm{if $i < M \land j < M \lor i \geq M  \land j \geq M $}\\
  \end{array} \right.
\end{equation}
where $\delta$ is the instrumental uncertainty associated with the electrometer range. Note that results of measurements within the same range are not statistically independent, since they are influenced by the same systematical error related to instrument calibration. On the other hand, it was considered that data from different ranges are statistically independent.  

\begin{figure}
\includegraphics[height=4.5cm, width = 8.7cm]{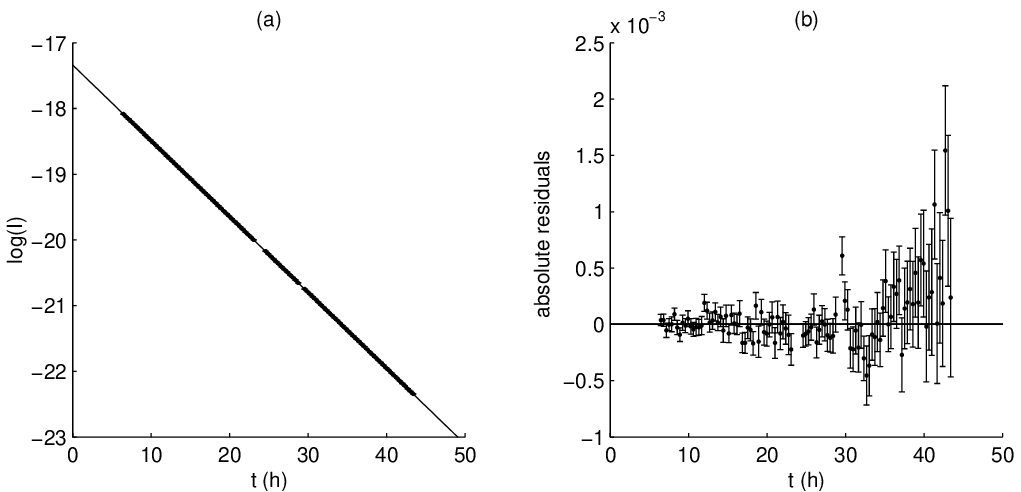}
\caption{\label{fig:graph2}  Linear least square fit of the logarithm of the ionization current as a function of time based on the model proposed and residuals of the experimental points and the fitted values.}
\end{figure}

The linear least square fit and the residuals are shown in Figure \ref{fig:graph2} and the results are shown in Table \ref{tab:table1}. Residuals of the order of $0.01\%$ were obtained, which indicates good linearity. Besides, the value of $P(\chi^{2})$ indicates a good fit. However, graphical analysis of residuals shows a tendence in measurements with lower activities. This effect might have been caused by saturation effects or radionuclidic impurities.

\section{Conclusion}
The ionization current measurements showed a non-linear behavior, which was proved to be mainly due to the electrometer range switching. The hypothesis that this non-linearity did not influence the decay constant was tested, showing that this non-linearity can be corrected by the method proposed. Good linearity was then obtained. The experimental half-life obtained was 6.00757(13)h, which agrees to within 0.01$\%$ with the value 6.0072(9)h  taken from the literature \cite{Tuli}.   

\begin{table}
\caption{\label{tab:table1} Least-square fitting parameters from $^{99m}$Tc decay. $\alpha_{1}$ and  $\alpha_{2}$($\sigma$) are the intercept, $t_{1/2}$ is the $^{99m}$Tc half-live,and $P(\chi^{2})$ is the probability of obtaining either $\chi^{2}$ or a larger value given f degrees of freedom.}
\begin{ruledtabular}
\begin{tabular}{ccccc} 
   $\alpha_{1}$($\sigma$) &  $\alpha_{2}$($\sigma$)   & $t_{1/2}$($\sigma$) & $P(\chi^{2}) \%$ & f \\ 
\hline	
   -17.337736(34)  &  -17.336369(80) & 6.00757(13) &    $\pm$ 39.2 & 101 \\
\end{tabular}
\end{ruledtabular}
\end{table}

\end{document}